\documentclass[12pt]{article}
\usepackage{ssi}
\usepackage{times}
\usepackage{graphicx,here,rotate,epsf,epsfig}

\begin{document}

\title{Cosmic Connections}

\author{John Ellis\thanks{
\noindent
\copyright\ 2003 by John Ellis. }\\
Theory Division \\
CERN, CH-1211 Geneva 23, Switzerland \\[0.4cm]
}

%
%
%
 
\newcommand{\mycomm}[1]{\hfill\break{ \tt===$>$ \bf #1}\hfill\break}

\def\ga{\mathrel{\raise.3ex\hbox{$>$\kern-.75em\lower1ex\hbox{$\sim$}}}}
\def\la{\mathrel{\raise.3ex\hbox{$<$\kern-.75em\lower1ex\hbox{$\sim$}}}}
\def\gev{{\rm \, Ge\kern-0.125em V}}
\def\tev{{\rm \, Te\kern-0.125em V}}
\def\beq{\begin{equation}}
\def\eeq{\end{equation}}
\def\st{\scriptstyle}
\def\ss{\scriptscriptstyle}
\def\mb{m_{\widetilde B}}
\def\msf{m_{\tilde f}}
\def\mst{m_{\tilde t}}
\def\mf{m_{\ss{f}}}
\def\mpar{m_{\ss\|}^2}
\def\mpl{M_{\rm Pl}}
\def\mchi{m_{\chi}}
\def\ohsq{\Omega_{\chi} h^2}
\def\msn{m_{\tilde\nu}}
\def\m12{m_{1\!/2}}
\def\mstpl{m_{\tilde t_{\ss 1}}^2}
\def\mstpr{m_{\tilde t_{\ss 2}}^2}

\def\sm{Standard Model}

\def\ga{\mathrel{\raise.3ex\hbox{$>$\kern-.75em\lower1ex\hbox{$\sim$}}}}
\def\la{\mathrel{\raise.3ex\hbox{$<$\kern-.75em\lower1ex\hbox{$\sim$}}}}
\def\gyr{{\rm \, G\kern-0.125em yr}}
\def\gev{{\rm \, Ge\kern-0.125em V}}
\def\tev{{\rm \, Te\kern-0.125em V}}
\def\ss{\scriptscriptstyle}
\def\scs{\scriptstyle}
\def\mb{m_{\widetilde B}}
\def\mst{m_{\tilde\tau_R}}
\def\mstop{m_{\tilde t_1}}
\def\msl{m_{\tilde{\ell}_1}}
\def\stau{\tilde \tau}
\def\stop{\tilde t}
\def\sbot{\tilde b}
\def\mchi{m_{\tilde \chi}}
\def\mxi{m_{\tilde{\chi}_i^0}}
\def\mxj{m_{\tilde{\chi}_j^0}}
\def\mchari{m_{\tilde{\chi}_i^+}}
\def\mcharj{m_{\tilde{\chi}_j^+}}
\def\mgluino{m_{\tilde g}}
\def\msf{m_{\tilde f}}
\def\m12{m_{1\!/2}}
\def\mtb{\overline{m}_{\ss t}}
\def\mbb{\overline{m}_{\ss b}}
\def\mfb{\overline{m}_{\ss f}}
\def\mf{m_{\ss{f}}}
\def\gt{\gamma_t}
\def\gb{\gamma_b}
\def\gf{\gamma_f}
\def\thm{\theta_\mu}
\def\tha{\theta_A}
\def\thb{\theta_B}
\def\mgl{m_{\ss \tilde g}}
\def\cp{C\!P}
\def\ch{{\widetilde \chi}} 
\def\st{{\widetilde \tau}_{\scriptscriptstyle\rm 1}}
\def\sel{{\widetilde e}_{\scriptscriptstyle\rm R}}
\def\sl{{\widetilde \ell}_{\scriptscriptstyle\rm R}}
\def\msn{m_{\ch}}
\def\tsq{|{\cal T}|^2}
\def\tcm{\theta_{\rm\scriptscriptstyle CM}}
\def\half{{\textstyle{1\over2}}}
\def\neq{n_{\rm eq}}
\def\qeq{q_{\rm eq}}
\def\slash#1{\rlap{\hbox{$\mskip 1 mu /$}}#1}%
\def\mw{m_W}
\def\mz{m_Z}
\def\mhb{m_{H}}
\def\mhl{m_{h}}
\newcommand\f[1]{f_#1}
\def\nl{\hfill\nonumber\\&&}

\def\gappeq{\mathrel{\rlap {\raise.5ex\hbox{$>$}}
{\lower.5ex\hbox{$\sim$}}}}

\def\lappeq{\mathrel{\rlap{\raise.5ex\hbox{$<$}}
{\lower.5ex\hbox{$\sim$}}}}

\def\Toprel#1\over#2{\mathrel{\mathop{#2}\limits^{#1}}}
\def\FF{\Toprel{\hbox{$\scriptscriptstyle(-)$}}\over{$\nu$}}

\newcommand{\Zee}{$Z^0$}


\def\Yi{\eta^{\ast}_{11} \left( \frac{y_{i}}{2} g' Z_{\chi 1} + 
        g T_{3i} Z_{\chi 2} \right) + \eta^{\ast}_{12} 
        \frac{g m_{q_{i}} Z_{\chi 5-i}}{2 m_{W} B_{i}}}

\def\Xii{\eta^{\ast}_{11} 
        \frac{g m_{q_{i}}Z_{\chi 5-i}^{\ast}}{2 m_{W} B_{i}} - 
        \eta_{12}^{\ast} e_{i} g' Z_{\chi 1}^{\ast}}

\def\Wi{\eta_{21}^{\ast}
        \frac{g m_{q_{i}}Z_{\chi 5-i}^{\ast}}{2 m_{W} B_{i}} -
        \eta_{22}^{\ast} e_{i} g' Z_{\chi 1}^{\ast}}

\def\Vi{\eta_{22}^{\ast} \frac{g m_{q_{i}} Z_{\chi 5-i}}{2 m_{W} B_{i}}
        + \eta_{21}^{\ast}\left( \frac{y_{i}}{2} g' Z_{\chi 1}
        + g T_{3i} Z_{\chi 2} \right)}

\def\zthree{\delta_{1i} [g Z_{\chi 2} - g' Z_{\chi 1}]}

\def\zfour{\delta_{2i} [g Z_{\chi 2} - g' Z_{\chi 1}]}


\begin{center}
{\it Closing Lecture at the 31st SLAC Summer Institute, August 2003: PSN 
TF07}\\
\end{center}
\begin{center}
CERN-TH/2003-270 $\; \; \; \;$ {\tt astro-ph/0310913}
\end{center}

\maketitle
\begin{abstract}%
\baselineskip 16pt
A National Research Council study on connecting quarks with the cosmos has
recently posed a number of the more important open questions at the
interface between particle physics and cosmology. These questions 
include the nature of dark matter and dark energy, how the Universe 
began, modifications to gravity, the effects of neutrinos on the 
Universe, how cosmic accelerators work, and whether there are new 
states of matter at high density and pressure. These questions are 
discussed in the context of the talks presented at this Summer Institute.
\end{abstract}

\section{Connecting Quarks with the Cosmos}

My task in this closing lecture is to preview possible future developments
at the interface between particle physics on one side, and astrophysics
and cosmology on the other side. Though these cosmic connections may
benefit from some theoretical advice, they must rely on the firm facts
provided by accelerator experiments, as well as non-accelerator
experiments and astronomical observations. To guide the discussion, I
structure this talk around a report with the same title as this section,
published recently by the U.S.  National Research Council~\cite{CQC}, that 
poses eleven major cosmological questions for the new century:

\noindent
$\bullet$ 1: {\it What is the dark matter?} \\
$\bullet$ 2: {\it What is the nature of dark energy?} \\
$\bullet$ 3: {\it How did the Universe begin?} \\
$\bullet$ 4: {\it Did Einstein have the last word on gravity?} \\
$\bullet$ 5: {\it What are the effects of neutrinos on the Universe?} \\
$\bullet$ 6: {\it How do cosmic accelerators work?} \\
$\bullet$ 7: {\it Are protons unstable?} \\
$\bullet$ 8: {\it Are there new states of matter at high density and 
pressure?} \\
$\bullet$ 9: {\it Are there additional space-time dimensions?} \\
$\bullet$ 10: {\it How were heavy elements formed?} \\
$\bullet$ 11: {\it Do we need a new theory of matter and light?} \\

The last two questions primarily concern nuclear physics and plasma
physics, respectively, and I do not discuss them here. A particle
physicist's answer to the fourth question about the completeness of
general relativity is inextricably linked to the ninth question about
extra dimensions. Likewise, the fifth and seventh questions about
neutrinos and protons, respectively, are linked in grand unified theories.  
Therefore, I treat these questions in pairs.

\section*{1: What is the dark matter?}

We have heard repeatedly at this institute that dark matter is necessary 
for the formation of structures in the 
Universe~\cite{Blandford,Dekel,Wittman}. 
The latest data from the 
Sloan Digital Sky Survey~\cite{SDSS}, shown here by Kent~\cite{Kent}, are 
very consistent with the 
power spectrum measured in the CMB and by previous sky surveys, weak 
lensing and the Lyman-$\alpha$ forest. At the level of galaxy clusters, as 
we heard here from Henry~\cite{Henry}, some resemble train wrecks and are 
still forming 
today, whereas others have relaxed and are good probes of the dark matter 
content. Even before the combination of Type-1a supernovae and the CMB, 
cluster data indicated that $\Omega_m \ll 1$: current cluster data 
yield~\cite{Henry}:
\begin{equation}
\Omega_m \; = \; 0.30^{+0.04}_{-0.03}
\label{clusterO}
\end{equation}
after marginalizing over $\Omega_b$ and $h$. Moreover, as discussed here 
by Dekel~\cite{Dekel}, the motion of luminous matter in the neighbourhood 
of our galaxy provides a detailed profile of the local dark matter density.

Is this dark matter composed of particles or of larger objects such as
white dwarfs or black holes? The recently-demonstrated concordance between
the values of $\Omega_b$ extracted from Big-Bang Nucleosynthesis and the
CMB~\cite{WMAP} confirms that the dark matter cannot be composed of
baryons, excluding a dominant white dwarf component and implying that any
substantial black hole component must have been primordial. Microlensing
searches exclude the possibility that our own galactic halo is composed of
objects weighing between $\lappeq 10^{-3}$ and $\gappeq 10^{+3}$ times the
mass of the Sun~\cite{MACHOs}.  Therefore, in the following, we
concentrate on particle candidates for the dark matter.

Is this dark matter hot, warm or cold? The recent WMAP~\cite{WMAP}, 
2dF~\cite{2dF} and SDSS data~\cite{SDSS} are
very consistent with the standard cold dark matter paradigm. In
particular, the combination of WMAP with other data implies that
\begin{equation}
\Omega_{HDM} h^2 \; < \; 0.0076,
\label{WMAPhot}
\end{equation}
corresponding to $\Sigma_\nu m_\nu < 0.7$~eV. Moreover, the early
reionization of the Universe recently discovered by WMAP~\cite{WMAPpol} 
requires some structures to have started forming very early, which is 
evidence against warm dark matter.

However, there are problems with the cold dark matter paradigm. For one
thing, the density profiles of galactic cores appear less singular than
calculated in some cold dark matter simulations~\cite{cores} - but these
may be changed by interactions with ordinary matter and by mergers and
black hole formation~\cite{Merritt}. For another thing, there is little
observational evidence for the halo substructures predicted by cold dark
matter simulations - but the formation of stars may be
dynamically inhibited in small structures near larger
galaxies. Therefore, we continue to focus on cold dark matter
candidates.

Generally speaking, these might have been produced by some thermal
mechanism in the very early Universe, or non-thermally. A good example of
the latter is the axion~\cite{axion}, which is my second-best candidate
for cold dark matter. Recent data from the LLNL axion search, reported
here by Nelson~\cite{Nelson}, excludes the possibility that a KSVZ axion
weighing between 1.9 and 3.4~$\mu$eV could constitute our galactic halo.

Another example of non-thermally produced cold dark matter could be a
superheavy particle produced around the epoch of inflation~\cite{Chung},
called by Kolb~\cite{Kolb} the `wimpzilla'. A natural example of a
wimpzilla is a metastable `crypton' from the hidden sector of some string
model~\cite{cryptons}. If metastable, a wimpzilla could be the orgin of
the ultra-high-energy cosmic rays discussed here by Ong~\cite{Ong}.

The classic thermally-produced cold dark matter candidate is the lightest
supersymmetric particle (LSP)~\cite{EHNOS}, but another possibility
proposed recently is the lightest Kaluza-Klein particle (LKP) in some
scenarios with universal extra bosonic dimensions (UED)~\cite{LKP}. The
spectra in some UED models are strikingly similar to those in
supersymmetric models, but with bosons and fermions switched around.

During this institute, there was an important update for the accelerator
constraints on supersymmetry, with a re-analysis of the $e^+ e^-$ data
used to estimate the Standard Model contribution to the anomalous
magnetic moment of the muon, $g_\mu - 2$~\cite{newDavier}. These now bring
the Standard Model prediction to within 2 $\sigma$ of the experimental
value, leaving less room for a supersymmetric contribution~\cite{Fred}.  

The direct searches for LSP dark matter were reviewed here by
Spooner~\cite{Spooner}. As he mentioned, the long-running DAMA claim to
have observed a possible annual modulation signal for cold dark matter
scattering has recently been reinforced by new data from the same
experiment that show the annual modulation persisting for seven
years~\cite{newDAMA}. However, several other experiments, including
CDMS~\cite{CDMS}, EDELWEISS~\cite{EDELWEISS} and most recently
ZEPLIN~1~\cite{Spooner} exclude a spin-independent scattering cross
section in the range proposed by DAMA. This range is also far above what
one calculates in the CMSSM when one takes into account all the
constraints~\cite{EFFMO}. More worryingly, the ICARUS
collaboration has recently measured a large annual modulation in the
neutron flux in the Gran Sasso laboratory where DAMA is
located~\cite{private}.

What are the prospects for detecting dark matter at a particle
accelerator? First at bat is the Fermilab Tevatron collider, which, as we
heard here from Thomson~\cite{Thomson}, now aims at an integrated
luminosity of 2~pb$^{-1}$ by 2007 and 4~pb$^{-1}$ by 2009. This will
enable it to search for squarks and gluinos with masses considerably
heavier than the present limits. Next at bat will be the LHC, which is
scheduled to start making collisions in 2007. With a centre-of-mass energy
of 14~TeV and a luminosity of $10^{34}$~cm$^{-2}$s$^{-1}$, it will be able
to find squarks and sleptons if they weigh $\lappeq 2.5$~TeV~\cite{LHC}.
If the squark and gluino masses are relatively low, measurements at the
LHC may fix the supersymmetric model parameters sufficiently accurately to
enable $\Omega_\chi h^2$ to be calculated with an accuracy comparable to
the uncertainty currently provided by WMAP~\cite{newBench}.  
The LHC will also address many
other issues of interest to cosmology, such as the origin of mass, which
may be linked to the mechanism for inflation, the primordial plasma in the
very early Universe, and the cosmological matter-antimatter asymmetry.

Most analyses of supersymmetric dark matter assume that the lightest
supersymmetric particle (LSP) is the lightest neutralino, a mixture of
spartners of Standard Model particles. However, another possibility,
discussed here by Feng~\cite{Feng}, is that the LSP is the supersymmetric
partner of the graviton, the gravitino~\cite{Fengetal}. This possibility
is severely constrained by the concordance between Big-Bang
nucleosynthesis and CMB~\cite{CEFO}.  However, the possibility remains of
a deviation from standard Big-Bang nucleosynthesis calculations and/or a
distortion of the CMB spectrum.

\section*{2: What is the nature of dark energy?}

The necessity of dark energy became generally accepted when data on
high-redshift supernovae were combined with the CMB data favouring
$\Omega_{tot} \simeq 1$~\cite{SN}. This conclusion has been supported by
recent data extending the previous supernova samples to larger redshift
$z$, in particular~\cite{newSN}, but how robust is this conclusion? As has
already been mentioned, the pre-existing data on dark matter in clusters
have long favoured $\Omega_{matter} \simeq 0.3$ which, combined with the
CMB data, favour dark energy $\Lambda$ with $\Omega_\Lambda \simeq 0.7$
independently from the supernova data~\cite{DM}. Moreover, as was
discussed here by Kolb~\cite{Kolb} and Pinto~\cite{Pinto}, there are good
reasons to think that the Type-1a supernovae are indeed good standard
candles. Also, as discussed here by Wright~\cite{Wright}, radical
alternatives to the standard $\Lambda$CDM scenario such as modified
Newtonian dynamics (MOND)~\cite{Mond} do not agree with the CMB data. So
it seems that we have to learn to live with dark energy. Supporting
evidence for dark energy comes from the recent observation of the
integrated Sachs-Wolfe effect~\cite{ISW}, a correlation between galaxy
clusters and features in the CMB that appears only if there is dark energy
causing the space between clusters to expand.

The next question is whether this dark energy is constant, or whether it
is varying with time. The latter option offers the hope of understanding
why the dark energy density in the Universe today is similar in magnitude
to the density of matter, through some sort of `tracker
solution'~\cite{Q}. In this case, the dark energy would have non-trivial
dynamics described by an equation of state that can be parametrized by
$w(z) \equiv p(z) / \rho(z)$, where I emphasize that $w(z)$ depends in
general on the redshift $z$. Discarding this possibility for the moment,
the present cosmological data favour $w \simeq -1$, corresponding to a
cosmological constant, as discussed here by Kolb~\cite{Kolb}. The SNAP
satellite project~\cite{SNAP} aims at increasing substantially the
available sample of high-$z$ supernovae, and offers the prospect of
constraining $w(z)$ much more tightly. This may enable a clear distinction
to be drawn between time-varying `quintessence' models and a cosmological
constant.

If the vacuum energy $\Lambda$ is indeed constant, the next step will be
to calculate it. This is surely the ultimate challenge for any pretender
for a full quantum theory of gravity, such as string/M theory. For some
time, the efforts of the string community were directed towards proving
that $\Lambda = 0$. However, this was never achieved, despite searches for
a suitable symmetry or dynamical relaxation mechanism. Presumably a
non-zero value of $\Lambda$ is linked to microphysical parameters such as
$m_W, m_t, m_{susy}, \Lambda_{QCD}$, etc., and the challenge is to find
the right formula~\footnote{String theorists are also worried that,
whether $\Lambda$ is constant or not, the existence of an event horizon
appears inevitable. In this case, it is never possible to make exact
predictions because of information loss across the horizon.}.

If, on the other hand, $\Lambda$ is really varying, the next question is:  
what is the asymptotic value? Is it zero, a non-zero constant, or even $-
\infty$? Quintessence only postpones the problem.

\section*{3: How did the Universe begin?}

By now, the standard answer to this question is: inflation~\cite{Turner}.  
But this answer is far from being established. Simple models predict a
near-scale-invariant spectrum of near-Gaussian perturbations with a
model-dependent ratio of tensor and scalar modes. Some of these
predictions are successful: for example, the spectral index of the scalar
perturbations seen so far is consistent with being scale-invariant, with
an accuracy of a few \% when WMAP data are combined with data on
large-scale structure~\cite{WMAPinf}. However, one can never `prove' that
a statistical distribution is Gaussian: one can apply various tests, but
if they are passed, one can never be sure that the distribution will not
fail some future test. And there are some puzzles in the
WMAP spectrum, for example glitches around $\ell \simeq 100, 200$ and
$340$, as discussed here by Wright~\cite{Wright}. As for the possible
tensor modes, the first CMB polarization measurements have been published
by DASI and WMAP, whose sensitivity is close to expectations in some
inflationary models, but far above some predictions, as discussed here by
Winstein~\cite{Winstein}.

Assuming the validity of the basic inflationary paradigm poses a new
series of questions. Was inflation driven by some simple field-theoretical
mechanism, such as an $m^2 \phi^2$ or $\lambda \phi^4$ potential, or was
some more subtle (quantum-gravitational? stringy?) mechanism responsible?  
a`string plasma'? The WMAP measurements strongly disfavour the simplest
$\lambda \phi^4$ potential~\cite{WMAPinf}, but the $m^2 \phi^2$ potential
survives for now~\cite{ERY}. If inflation was driven by some scalar
inflaton $\phi$, how can it be related to the rest of particle physics?
The most suitable candidate in the present particle menagerie appears to
me to be the supersymmetric partner of the heavy neutrino in a seesaw
model of light neutrino masses~\cite{ERY}.  Even if inflation was
driven by a scalar inflaton field, the CMB might reveal some traces of
Planckian physics in the form of some effects suppressed by powers of
$m_P$.

However, to answer the question in the title of this section, one must
look beyond inflation, which presumably occurred when the energy density
in the Universe was ($\sim 10^{16}$~GeV)$^4$, back to when it approached
the Planck energy density ($\sim 10^{19}$~GeV)$^4$. At this epoch, perhaps
the Universe was described by some form of string cosmology or
pre-Big-Bang scenario~\cite{preBB}. How to test such an idea? One
possibility might be provided by gravitational waves~\cite{Landry} from
this epoch.

\section*{4/9: Does completing Einstein's theory of gravity require extra 
dimensions?}

Einstein certainly did not have the last word on gravity. His General
Theory of Relativity was one of the greatest physics achievements of the
first half of the twentieth century, the other being Quantum Mechanics.  
Combining them into a true quantum theory of gravity was the greatest
pieces of unfinished business of twentieth-century physics: in particular,
how to make sense of the uncontrollable infinities encountered when
gravitational interactions are treated perturbatively, and how to deal
with the loss of information apparently inherent in non-perturbative
gravitational phenomena such as black holes? Presumably the answers to
these questions involve modifying either General Relativity, or Quantum
Mechanics, or both.

The best/only candidate we have for a quantum theory of gravity is
string/M theory, which relies heavily on the existence of extra
dimensions.  These include fermionic dimensions, in the form of
supersymmetry with its accompanying superspace~\cite{susy}, as well as
`conventional' extra bosonic dimensions~\cite{JMR}. If they are to aid in
stabilizing the mass hierarchy, provide the cold dark matter and
facilitate unification of the particle interactions, the fermionic
dimensions should appear at the TeV scale, within reach of
colliders~\cite{hierarchy}. But what might be the scales of the extra
bosonic dimensions? Consistency of string theory at the quantum level
requires extra dimensions at the scale of $\sim 10^{33}$~cm, and
unification of gravity with the other interactions suggests they might
appear at $\sim 10^{29}$~cm~\cite{HW}. Colliders can probe distance scales
down to $\sim 10^{17}$~cm, but there is no particular reason to expect
that extra dimensions will show up at such a large scale.

What other signatures might there be for a quantum theory of gravity? One
possibility might be gravitational waves, or there might be signatures in
the CMB, as discussed earlier. It could even be that the inflation now
being probed by the CMB was produced by some stringy effect.  
As also discussed earlier, the value of the vacuum energy should be
calculable in a complete quantum theory of gravity. Other possible tests
of models of quantum gravity include the propagation of energetic
particles - which might be retarded by space-time foam, as could be probed
by measurements of photons from AGNs or GRBs~\cite{AEMNS} - or their
interactions, as could be probed by UHECRs. Modifications of
quantum mechanics could be probed by laboratory studies of K mesons, B
mesons and neutrons~\cite{EHNS}. There are plenty of ways in which
theories completing Einstein's theory of gravity can be tested.

\section*{5/7:  What are the effects of GUTs on the Universe?}

The direct upper limits on neutrino masses: $m_{\nu_e} \lappeq 2.5$~eV, 
$m_{\nu_\mu} \lappeq 190$~keV, $m_{\nu_\tau} \lappeq 18$~MeV~\cite{PDG}, 
have left 
open the possibility that neutrinos might be an important contribution to 
the dark matter. However, the combination of WMAP data with previous 
astrophysical and cosmological data provides the more stringent upper 
limit~\cite{WMAP}:
\begin{equation}
\Sigma_\nu m_\nu \; < \; 0.7~{\rm eV} \; \leftrightarrow \; 
\Omega_\nu h^2 \; < \; 0.0076,
\label{omeganu}
\end{equation}
implying that neutrinos can provide only a small fraction of the dark 
matter.

On the other hand, neutrino oscillation experiments tell us that neutrinos 
do have masses and mix~\cite{SK,SNO}. The minimal renormalizable model of 
neutrino 
masses requires the introduction of weak-singlet `right-handed' neutrinos 
$N$. These will in
general couple to the conventional weak-doublet left-handed neutrinos via
Yukawa couplings $Y_\nu$ that yield Dirac masses $m_D = Y_\nu \langle
0 \vert H \vert 0 \rangle \sim m_W$. In
addition, these `right-handed' neutrinos $N$ can couple to themselves via
Majorana masses $M$ that may be $\gg m_W$, since they do not require
electroweak summetry breaking. Combining the two types of
mass term, one obtains the seesaw mass matrix~\cite{seesaw}:
\begin{eqnarray}
\left( \nu_L, N\right) \left(
\begin{array}{cc}
0 & M_{D}\\   
M_{D}^{T} & M
\end{array}
\right)
\left(
\begin{array}{c}
\nu_L \\
N
\end{array}
\right),
\label{seesaw}
\end{eqnarray}
where each of the entries should be understood as a matrix in generation
space. The Dirac masses $M_{D}$ and large singlet-neutrino masses $M$ 
arise naturally in GUTs, but could appear even without all the GUT 
superstructure such as new gauge interactions.

The low-mass eigenstates resulting from the diagonalization of 
(\ref{seesaw}) do not, in general, coincide with flavour eigenstates, 
leading to neutrino oscillations described by the matrix
\begin{eqnarray}
V \; = \; \left(
\begin{array}{ccc}
c_{12} & s_{12} & 0 \\
- s_{12} & c_{12} & 0 \\
0 & 0 & 1   
\end{array}
\right)
\left(
\begin{array}{ccc}
1 & 0 & 0 \\
0 & c_{23} & s_{23} \\
0 & - s_{23} & c_{23}
\end{array}
\right)
\left(
\begin{array}{ccc}
c_{13} & 0 & s_{13} \\
0 & 1 & 0 \\
- s_{13} e^{- i \delta} & 0 & c_{13} e^{- i \delta}
\end{array}
\right).
\label{MNSmatrix}
\end{eqnarray}
Atmospheric neutrino oscillation experiments have established that $\sin^2
2 \theta_{23} \sim 1$ with $\Delta m^2 \sim 2.5 \times
10^{-3}$~eV$^2$~\cite{SK}, which is also consistent with data from the K2K
experiment~\cite{K2K}.  Solar neutrino experiments, particularly
SuperKamiokande~\cite{SKsun} and SNO~\cite{SNO}, have established that
$\tan^2 \theta_{12} \sim 0.5$ with $\Delta m^2 \sim 7 \times
10^{-5}$~eV$^2$, which is also consistent with data from the KamLAND
experiment~\cite{KamLAND}. On the other hand, we have only an upper limit
on the third mixing angle $\theta_{13}$, and no information on the
CP-violating phase $\delta$ in (\ref{MNSmatrix}).

The phase $\delta$ could in principle be measured by comparing the
oscillation probabilities for neutrinos and antineutrinos and computing 
the CP-violating asymmetry~\cite{DGH}:
\begin{eqnarray}
P \left( \nu_e \to \nu_\mu \right) - P \left( {\bar \nu}_e \to   
{\bar \nu}_\mu \right) \; & & = \;
16 s_{12} c_{12} s_{13} c^2_{13} s_{23} c_{23} \sin \delta \\ \nonumber
& & \sin \left( {\Delta m_{12}^2 \over 4 E} L \right)
\sin \left( {\Delta m_{13}^2 \over 4 E} L \right)
\sin \left( {\Delta m_{23}^2 \over 4 E} L \right),
\label{CPV}
\end{eqnarray}
using an intense neutrino super-beam, a $\beta$-beam or a neutrino 
factory~\cite{nufactYB}.

What does all this have to do with the Universe? In total, the minimal
seesaw model outlined above has 18 parameters, 9 of which are observable
at low energies: 3 light neutrino masses, 3 real mixing angles and 3
CP-violating phases (the oscillation phase $\delta$ and two others that
appear in neutrinoless double-$\beta$ decay. The other 9 parameters are 
associated with the heavy neutrino sector, and comprise 3 more masses, 3 
more real mixing angles and 3 more CP-violating phases.
CP violation in the neutrino sector offers the possibility of 
generating the baryon asymmetry of the Universe via heavy neutrino 
decays~\cite{FY}, which could generate a lepton asymmetry via
\begin{equation}
\Gamma ( N \to \ell + H ) \; \ne \; \Gamma ( N \to {\bar \ell} + H ).
\label{leptonas}
\end{equation}
Non-perturbative electroweak interactions would then transform part of 
this lepton asymmetry into the required baryon asymmetry.

The question then arises whether this baryon asymmetry is directly related
to the CP-violating parameter $\delta$ that could be observed in neutrino
oscillations. Unfortunately, the answer is no in general~\cite{ER}. Except
in specific models~\cite{FG}, this leptogenesis mechanism is independent
of the mixing angles and phases in the light neutrino sector. However,
neutrino oscillation experiments can demonstrate the principles on which
the leptogenesis mechanism is based.

There is another role that neutrino physics might have played in the early
Universe: the inflaton could have been a heavy 
sneutrino~\cite{ERY}. The WMAP data on
the scalar spectral index, the tensor/scalar ratio, etc., are consistent
with a simple $m^2 \phi^2$ model for inflation with $m \simeq 2 \times
10^{13}$~GeV~\cite{ERY}. This is comfortably within the range favoured by 
the seesaw
model for a heavy (s)neutrino. Moreover, if inflation was driven by a 
sneutrino, leptogenesis would have followed automatically. In order to 
avoid over-producing gravitinos following inflation, the reheating 
temperature should not exceed a few $\times 10^7$~GeV, which constrains 
the inflaton sneutrino couplings as well as its mass. 

How may one test such a scenario? Accelerators may play a role, since the
sneutrino inflaton model makes some relatively precise predictions for
processes violating charged-lepton number conservation~\cite{ERY}. 
These are quite close to the present
experimental upper limits, and we heard from Aihara~\cite{Aihara} of a new
upper limit $B(\tau \to \mu \gamma) < 3.2 \times 10^{-7}$, with the
prospect of further improvement as the B factories gather more data.

As we heard here from Prell~\cite{Prell}, the B factory measurements of CP
violation in $B^0 \to J/\psi K^0$ decays agree well with the Standard
Model: $\sin 2 \beta = 0.731 \pm 0.056$. On the other hand, data on $B^0
\to \phi K^0$ and other decays dominated by $b \to s$ penguin diagrams do
not agree so well with each other or with the Standard Model~\cite{phiKs}.
This is a place where the first deviations from the Standard Model of CP
violation are expected in some scenarios, such as GUT extensions of the
seesaw model~\cite{HS}.  However, any such effect must respect the
constraint imposed by the electric dipole moment of $^{199}$Hg~\cite{HS2}.
If a deviation does get confirmed by future data from B factories, CP
violation in the quark sector might be reinstated as a candidate for
baryogenesis, a role that the Standard Model of CP violation cannot play.

\section*{6: How do cosmic accelerators work?}

Candidates for the origins of the cosmic rays include neutron stars, white
dwarfs, supernova remnants, AGNs, GRBs, colliding galaxies and more, where
the first two might be responsible for the lower-energy cosmic rays
believed to originate within our galaxy, and the latter might be
responsible for the higher-energy cosmic rays believed to come from
outside our galaxy, as discussed here by Ong~\cite{Ong}.

As we heard here from Kahn~\cite{Kahn}, the cosmic X-ray background is now
getting much better understood, and seems to be largely due to discrete
sources.  As reported here by Teegarden~\cite{Teegarden}, the INTEGRAL
satellite has recently discovered a new class of X-ray sources for us to
understand, even if they do not contribute to the cosmic rays.

Progress was reported here in observations of some of the prospective
cosmic-ray sources. As discussed by Tanimori~\cite{Tanimori}, CANGAROO
observations of photons from RXJ1713.7-3946 provide eveidence of $\pi^0$
production by accelerated protons. Meanwhile observations of AGNs indicate
that they are probably powered by the accretion of matter onto black holes
weighing $10^6 - 10^9$ solar masses, which produce jets of relativistic
outflow. There is no strong evidence yet that they contribute to the
observed cosmic-ray spectrum, though a correlation between the arrival
directions of UHECRs and BL-Lac objects has been claimed~\cite{Lausanne}.

As for GRBs, the evidence is strengthening that (at least some of) the
longer-duration GRBs are associated with supernovae at high redshifts $z
\sim 1$. However, the shorter-duration GRBs with harder spectra may have
different origins.

We can expect light to be cast on $\gamma$-ray sources such as galactic
objects, AGNs, and GRBs by the GLAST satellite~\cite{GLAST}, which should
see $\sim 10^4$ sources each year. One of the interesting places to look
for $\gamma$-ray emission is the core of the galaxy. In some models, the
annihilations of supersymmetric relic particles in the core of the galaxy
would produce $\gamma$ rays detectable by GLAST~\cite{EFFMO}.

The origin of the UHECRs with energies $\gappeq 10^{20}$~GeV remains an
enigma. One would have expected these to be cut off by photo-absorption on
CMB photons~\cite{GZK}, so their observation by AGASA came as a 
suprise~\cite{AGASA}.  
However, more recently HiRes~\cite{HiRes} has filed to reproduce the AGASA
data, so it is unclear whether there is any excess to explain. Perhaps
they are just the tail of conventional cosmic rays produced by some
bottom-up mechanism.  Alternatively, perhaps they are produced by the
decays of netastable ultra-massive particles~\cite{Berez}. As discussed
here by Kolb~\cite{Kolb}, these might have been produced non-thermally
around the epoch of inflation. There are stringy models with suitably
long-lived metastable particles - `cryptons'~\cite{cryptons}, and
simulations of such a top-down decay mechanism are consistent with the
UHECR spectrum reported by AGASA~\cite{SS}. It will be exciting to see
whether the Auger project~\cite{Auger} now starting to take data in
Argentina is able to reproduce the AGASA data.

In addition to photons and protons, one might also expect the Cosmos to
send energetic neutrinos in our direction. Various projects to look for
these - AMANDA, ANTARES, NESTOR and IceCube - are underway, and will have
sufficient sensitivity to see some of the postulated extragalactic
sources. They may also be able to see the neutrino produced by the
photo-absorption reaction $p + \gamma \to n + (\pi^+ \to \nu)$ on the CMB. 
These detectors might also be able to observe energetic neutrinos from the 
annihilations of supersymmetric relic particles inside the Sun or 
Earth~\cite{EFFMO}.

\section*{8: Are there new states of matter at high density and pressure?}

Heading back towards the very early Universe, the first new state of
matter that we expect to encounter is the quark-gluon plasma. Lattice
gauge-theory simulations indicate a transition to this phase when the
temperature exceeds about 170~MeV~\cite{QGP}, which would have been the
case when the Universde was less than a few $\times 10^{-6}$~s old. There
is recurrent speculation that the transition to hadronic matter might
generate inhomogeneities with observable consequences. There have also
been conjectures that the cores of (at least some) neutron stars might be
made of quark-gluon matter, which could have implications for
core-collapse supernovae, neutron-star mergers and GRBs.

Accelerator experiments, first at CERN and more recently at the BNL RHIC
heavy-ion collider, produce in the laboratory dense and hot conditions
under which the quark-gluon transition should occur. There have been
tantalizing hints of quark-gluon matter, such as enhanced abundances of
strange particles and the suppression of $J/\psi$
production. As we heard here from Gagliardi~\cite{Gagliardi},
the RHIC experiments have recently observed an exciting new effect that
points towards the quark-gluon plasma. In high-energy proton-proton
or-deuteron collisions, the production of a high-$p_T$ jet due to hard
parton-parton scattering is accompanied by another jet in the opposite
azimuthal direction. However, this balancing jet is absent in central
Au-Au collisions~\cite{nojet}. The quark-gluon plasma interpretation is
that the parton that should have produced the opposite jet was quenched by
scattering on the naked quarks and gluons in the plasma, dispersing its
transverse energy. However, although this interpretation looks very
natural, it cannot yet be regarded as established.

Heading further back to when the temperature of the Universe exceeded
about 100~GeV and its age was $< 10^{-10}$~s, we believe that the Universe
was dominated by an electroweak plasma in which the Higgs mechanism was
switched off and Standard Model particles lost their masses. This picture
is supported by lattice simulations, but only experiments at the LHC will
provide us with all the information we need to calculate this phase
transition reliably.  If this electroweak phase transition was first
order, it would have provided an opportunity for electroweak baryogenesis,
an alternative to the leptogenesis scenario discussed earlier~\cite{EWPT}.

Heading even further back, in GUTs there could have been an analogous
phase transition in the very early Universe when it was $< 10^{-30}$~s
old. However, it is unclear what experimental signatures this might have
produced. Also in the very early Universe, there may have been a
transition to a `string plasma' phase. perhaps this is what laid down the 
perturbations seen in the CMB?

\section*{The most important question of all ...}

... is undoubtedly the one we have not yet had the `branes' to ask.

\end{document}